# An $L_0$-Norm Constrained Non-Negative Matrix Factorization Algorithm for the Simultaneous Disaggregation of Fixed and Shiftable Loads

Ahmad Khaled Zarabie
Electrical & Computer Engineering Dept.
Kansas State University
Manhattan, KS 66506, USA

Sanjoy Das
Electrical & Computer Engineering Dept.
Kansas State University
Manhattan, KS 66506, USA

*Abstract*—Energy disaggregation refers to the decomposition of energy use time series data into its constituent loads. This paper decomposes daily use data of a household unit into fixed loads and one or more classes of shiftable loads. The latter are characterized by ON/OFF duty cycles. A novel algorithm based on non-negative matrix factorization (NMF) for energy disaggregation is proposed, where fixed loads are represented in terms of real-valued basis vectors, whereas shiftable loads are divided into binary signals. This binary decomposition approach directly applies $L_0$-norm constraints on individual shiftable loads. The new approach obviates the need for more computationally intensive methods (e.g. spectral decomposition or mean-field annealing) that have been used in earlier research for these constraints. A probabilistic framework for the proposed approach has been addressed. The proposed approach's effectiveness has been demonstrated with real consumer energy data.

## I. INTRODUCTION

Energy disaggregation refers to the decomposition of energy usage into multiple components in a physically meaningful way [1],[2]. For instance, daily energy consumption of a household unit can be disaggregated into various loads, such as refrigerator, air conditioner, lighting, pool pump, and other loads that are typically present in homes. In disaggregation tasks, the data samples are of the form of energy use over a fixed duration of time, at regularly spaced intervals. Until the last decade, energy disaggregation had met with little success. However, in recent years, NMF [3] has emerged as powerful tool for this purpose and has met with remarkable success.

The classical NMF algorithm decomposes an input data matrix **X** whose columns are $D \times 1$ sample vectors, into two factors, **W** and **H**, so that their product equals **X**. Usually, **X** has a very large number of columns, which are independent samples. Although there exists an abundance of classical matrix methods to factorize a given matrix in such a manner, in NMF there is the additional constraint that $\mathbf{W} \geq \mathbf{0}$ and $\mathbf{H} \geq \mathbf{0}$ [4],[5]. This non-negativity requirement placed on both **W** and **H** render NMF suitable for many applications. For instance, NMF can be used to decompose image sequences or audio power spectra into factors **W** and **H** that can only have non-negative values, since neither pixel values nor power components can be negative. This is also the situation in the present research. The columns of **W**, which are relatively lesser in number, serve as basis vectors so that each sample $\mathbf{x}(n), n \in \mathcal{N}$ which are columns of **X** can be represented as a weighted combination of the bases, with the non-negative weights being the corresponding column vectors $\mathbf{h}(n)$ of **H**. In load disaggregation, the basis vectors may correspond to individual appliances [1],[2].

Due to the non-negativity constraints, **X** is not exactly factorizable into **W** and **H**; whence the goal of NMF is to seek an approximate solution, so that $\mathbf{X} \approx \mathbf{WH}$. NMF is an ill-posed problem since if $\mathbf{X} \approx \mathbf{WH}$ is one solution, then so are other factorizations of the form $\mathbf{X} \approx \mathbf{WPP^T H}$ where **P** is any rotation matrix that preserves non-negativity, that can be considered to be other solutions.

Furthermore, NMF involves seeking an optimum of a non-convex objective function. For example, although the Frobenius norm, $\|\cdot\|_F$, is convex in a single argument, this is not the case with NMF, where the objective function is $\|\mathbf{X} - \mathbf{WH}\|_F^2$, the square normed difference between **X** and **WH** (a very popular choice in many applications). This is due to the presence of the matrix product **WH** in the argument. Two other very commonly used objectives are the (generalized) Kullback-Leibler divergence and the Itakura-Saito divergence [6]. The latter is a special case of the $\beta$-divergence that is also used in some NMF applications [7]. From the non-convexity of the objective function, it follows that the NMF problem would encounter local minima. Although most NMF algorithms are first-order descent methods, providing local minima as the outputs, it has been seen that NMF consistently produces good results for a wide variety of applications.

Several approaches for NMF have been proposed, the earliest of them being the multiplicative method [3],[4]. Further details of this method are postponed until the following section. Another successful method is the alternating least squares algorithm [8]. In this approach, the KKT conditions at the optimum provide a set of equations through which iterative updates on the matrices **W** and **H** are carried out. Each iteration typically involves matrix pseudoinversion that is accomplished indirectly through LU decomposition. The projected gradient descent algorithm has also been applied for NMF [9]. This is a straightforward extension of gradient descent, with the matrices **W** and **H** thresholded at the end of each step to non-negative

---

This work was partially supported by the US National Science Foundation-CPS under Grant CNS-1544705.

values. More recent NMF algorithms include the hierarchical NMF [10] that is designed to deal with unduly large sized matrices, and the randomized NMF algorithm [11].

Earlier NMF algorithms that restrict either **W** or **H** to be sparse matrices tended to replace the non-differentiable $L_0$ norm constraint with an approximate measure of sparsity based on the $L_1$ and $L_2$ norms [12]. More recently, algorithms that make explicit use of the $L_0$ norm have also been suggested [13]. The algorithm in [13] applies *k*-SVD [14] based rank-1 updates on either of the two matrices. Therefore, its use is confined to Frobenius norm NMF.

The approach in [15] applies $L_0$ norm constraints (referred to as *sum-to-k* constraints in [15]) indirectly by including a penalty term to a Frobenius normed objective. It has been used for energy disaggregation of HVAC load components in an industrial building and in a smart home setting. Another method has been proposed in [2],[16] to impose $L_0$ constraints, which uses a softmax distribution for the elements in **H** to assign weights to them in such a manner that those with higher values are likelier to improve the objective function. Semi-supervised NMF using prior knowledge of the usage time profiles of individual appliances has been proposed in [17]. NMF has been applied for data over larger periods in time to glean seasonal trends in usage profiles in [18].

In this research, a novel NMF algorithm for load disaggregation has been proposed. It uses the Frobenius norm as objective, although the approach is generalized enough to be extended to others. The novelty of this approach is the manner in which the load is divided into two classes, (*i*) *fixed loads*, and, (*ii*) *shiftable loads*. This is a fundamental distinction between the different appliances in a typical household that has not been hitherto considered.

Fixed loads are associated with appliances that are in continual use throughout the day. Examples of shiftable loads include lighting, air-conditioners, and refrigerators. On the other hand, shiftable loads pertain to appliances that are used intermittently, such as washers, dryers, and ovens. The latter class of loads are characterized by duty cycles, with typical temporal profiles. Our approach tries to exploit this feature in shiftable loads to obtain improved load disaggregation. Moreover, direct $L_0$ norm constraints are imposed on the number of duty cycles, separately for each shiftable load.

## II. PROBLEM STATEMENT

### A. Load Models

Each input $\mathbf{x}(n)$ is a vector of $D$ regularly spaced samples of energy usage of a single household unit over a 24-hour period. Therefore, the index $n$ can be regarded as the energy use of the $n^{\text{th}}$ day in the sample set $\mathcal{N}$. The purpose of the proposed NMF algorithm is to express each sample in the following manner.

$$\mathbf{x}(n) \approx \sum_{k \in \mathcal{F}} h_k^f(n) \mathbf{w}_k^f + \sum_j \sum_{k \in \mathcal{S}_j} h_{j,k}^s(n) \mathbf{w}_{j,k}^s. \quad (1)$$

This work was partially supported by the US National Science Foundation-CPS under Grant CNS-1544705.

In (1), $\mathcal{F}$ is the fixed load basis set and $\mathcal{S}_j$ is that of the $j^{\text{th}}$ shiftable load. Each $D \times 1$ vector $\mathbf{w}_k^f$ is a fixed load basis vector $k$ ($k \in \mathcal{F}$). Likewise, each $D \times 1$ $\mathbf{w}_{j,k}^s$ is a basis vector $k$ ($k \in \mathcal{S}_j$) of the $j^{\text{th}}$ shiftable load. In terms of basis matrices,

$$\mathbf{X} \approx \widetilde{\mathbf{X}} = \mathbf{W}^f \mathbf{H}^f + \sum_j \mathbf{W}_j^s \mathbf{H}_j^s \quad (2)$$

The quantities $\mathbf{W}^f$ and $\mathbf{W}_j^s$ are basis matrices of dimensionalities $D \times |\mathcal{F}|$ and $D \times |\mathcal{S}_j|$. The matrices $\mathbf{H}^f = [\mathbf{h}^f(n)]_{n \in \mathcal{N}}$ and $\mathbf{H}_j^s = [\mathbf{h}_j^s(n)]_{n \in \mathcal{N}}$ are $|\mathcal{F}| \times N$ and $|\mathcal{S}_j| \times N$ dimensional arrays of weights $h_k^f(n)$ and $h_{j,k}^s(n) \in \{0,1\}$. This decomposition is illustrated in Fig. 1.

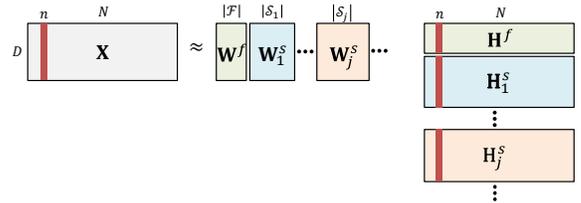

**Fig. 1**. Schematic of NMF factorization with separate basis and weight matrices for the fixed and shiftable loads.

cycles. It is assumed that during ON intervals, the $j^{\text{th}}$ shiftable load draws a constant amount of energy $p_j$ and that it can stay ON for a maximum of $L_j$ time intervals. Hence,

$$\left\| \mathbf{h}_j^s(n) \right\|_0 \leq L_j. \quad (3)$$

Fig. 2 shows the time profile of a shiftable load for the duration of a single day. It should be noted that as $L_j$ is the upper limit on the number of duty cycles, the actual number of such cycles may be less than $L_j$.

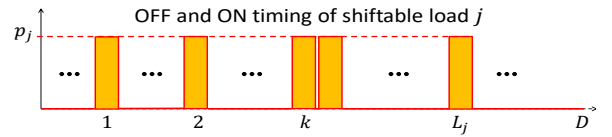

**Fig. 2**. Illustrative usage profile of a shiftable load showing a maximum of $L_j$ duty cycles, with each such cycle being a rectangular pulse of uniform peak $p_j$.

### B. Objective

The squared Frobenius norm of the difference between the real data matrix **X** and its approximation $\widetilde{\mathbf{X}}$,

$$\Phi(\widetilde{\mathbf{X}}|\mathbf{X}) = \frac{1}{2} \left\| \mathbf{X} - \widetilde{\mathbf{X}} \right\|_F^2$$
$$= \frac{1}{2} \sum_n \| \mathbf{x}(n) - \widetilde{\mathbf{x}}(n) \|_2^2. \quad (4)$$

The expression for the approximation $\widetilde{\mathbf{x}}(n)$ in (4) has already been provided in (1). The summation in (4) is carried out over all samples in the set $\mathcal{N}$. A probabilistic justification for this choice is provided in Section III, which also serves as a theoretical basis for obtaining the weights associated with the shiftable loads in the proposed approach.

III. APPROACH

## A. Probabilistic Framework

Consider the following probabilistic interpretation. The joint probability of $\mathbf{X}$, assuming that all samples are statistically *independent* is given by the following expression,

$$p[\mathbf{X}|\mathbf{W}^f, \mathbf{H}^f, \mathbf{W}^s_j, \mathbf{H}^s_j] = \prod_n p[\mathbf{x}(n)|\mathbf{W}^f, \mathbf{h}^f(n), \mathbf{W}^s_j, \mathbf{h}^s_j(n)]. \quad (5)$$

It is assumed that each sample $\mathbf{x}(n)$ follows a Gaussian distribution around its expected value $\tilde{\mathbf{x}}(n)$. In this case, the probability of each such sample can be expressed in the following manner,

$$p[\mathbf{x}(n)|\mathbf{W}^f, \mathbf{h}^f(n), \mathbf{W}^s_j, \mathbf{h}^s_j(n)]$$
$$= \frac{1}{\sigma^D (2\pi)^{\frac{D}{2}}} \prod_d e^{-\frac{1}{2\sigma^2}(x_d(n) - \tilde{x}_d(n))^2}. \quad (6)$$

The negated log probability of $\mathbf{x}(n)$ is,

$$-\log p[\mathbf{x}(n)|\mathbf{W}^f, \mathbf{h}^f(n), \mathbf{W}^s_j, \mathbf{h}^s_j(n)]$$
$$= \frac{1}{2\sigma^2} \sum_d (x_d(n) - \tilde{x}_d(n))^2 + \log \sigma^D (2\pi)^{\frac{D}{2}}$$
$$= \frac{1}{2\sigma^2} \|\mathbf{x}(n) - \tilde{\mathbf{x}}(n)\|_2^2 + \log \sigma^D (2\pi)^{\frac{D}{2}}. \quad (7)$$

Hence the negative log probability of $\mathbf{X}$ is,

$$-\log p[\mathbf{X}|\mathbf{W}^f, \mathbf{H}^f, \mathbf{W}^s_j, \mathbf{H}^s_j]$$
$$= \prod_n p[\mathbf{x}(n)|\mathbf{W}^f, \mathbf{h}^f(n), \mathbf{W}^s_j, \mathbf{h}^s_j(n)]$$
$$= \frac{1}{2\sigma^2} \|\mathbf{X} - \tilde{\mathbf{X}}\|_F^2 + N \log \sigma^D (2\pi)^{\frac{D}{2}}, \quad (8)$$

where the Frobenius norm of the difference between $\mathbf{X}$ and $\tilde{\mathbf{X}}$ is given by the earlier expression in (4).

Applying the maximum likelihood criterion to (8),

$$[\mathbf{W}^f, \mathbf{H}^f, \mathbf{H}^s_j]_{ML} = \underset{\mathbf{W}^f, \mathbf{H}^f, \mathbf{H}^s_j}{\arg\inf} \frac{1}{2} \|\mathbf{X} - \tilde{\mathbf{X}}\|_F^2. \quad (9)$$

The expression in (9) above provides a theoretical justification for the choice of objective function in (4) in the present approach.

The component of the objective function associated with the $n^{\text{th}}$ sample is,

$$\varphi(\tilde{\mathbf{x}}(n)|\mathbf{x}(n)) \triangleq \varphi(n)$$
$$= \frac{1}{2} \|\mathbf{x}(n) - \tilde{\mathbf{x}}(n)\|_2^2$$
$$= \frac{1}{2} \sum_d (x_d(n) - \tilde{x}_d(n))^2. \quad (10)$$

## B. Multiplicative Updates

Multiplicative updates are used for the matrices $\mathbf{W}^f$ and $\mathbf{H}^f$ that are associated with the fixed load. Although this method is quite routine in the existing literature on NMF, it is described here for the sake of completeness of this paper. More details can be found in [2],[3],[4],[5].

Consider any parameter $\mathbf{P}$ (which can be either $\mathbf{W}^f$ or $\mathbf{H}^f$). Let $\nabla_{\mathbf{P}} \Phi$ be the gradient of tan arbitrary objective function $\Phi$. Its gradient can be expressed in terms of its positive and negative components, so that $\nabla_{\mathbf{P}} \Phi = \nabla_{\mathbf{P}}^+ - \nabla_{\mathbf{P}}^-$. In gradient descent, the update rule would have been of the form,

$$\mathbf{P} \leftarrow \mathbf{P} - \eta \nabla_{\mathbf{P}}^+ + \eta \nabla_{\mathbf{P}}^-.$$

In the multiplicative method, the following multiplicative update replaces the gradient ascent step,

$$\mathbf{P} \leftarrow \mathbf{P} \circ \nabla_{\mathbf{P}}^- \oslash \nabla_{\mathbf{P}}^+.$$

In the proposed approach, the parameters $\mathbf{W}^f$ and $\mathbf{H}^f$ are subject to multiplicative updates. In order to do so, the derivatives of the objective must be first computed. It can be shown that,

$$\nabla_{\mathbf{W}^f} \frac{1}{2} \|\mathbf{X} - \tilde{\mathbf{X}}\|_F^2 =$$
$$-\mathbf{X} \circ \left( \mathbf{W}^f \mathbf{H}^f + \sum_j \mathbf{W}^s_j \mathbf{H}^s_j \right)^{\circ -1} \mathbf{H}^{f\text{T}} + \mathbf{1}_{D \times N} \mathbf{H}^{f\text{T}}. \quad (11)$$

$$\nabla_{\mathbf{H}^f} \frac{1}{2} \|\mathbf{X} - \tilde{\mathbf{X}}\|_F^2 =$$
$$-\mathbf{W}^{f\text{T}} \mathbf{X} \circ \left( \mathbf{W}^f \mathbf{H}^f + \sum_j \mathbf{W}^s_j \mathbf{H}^s_j \right)^{\circ -1} + \mathbf{W}^{f\text{T}} \mathbf{1}_{D \times N}. \quad (12)$$

Using the above expressions for the gradients, the update rules for the matrices pertaining to the fixed loads are,

$$\mathbf{W}^f \leftarrow \mathbf{W}^f \circ \frac{(\mathbf{X} \circ \tilde{\mathbf{X}}^{\circ -1}) \mathbf{H}^{f\text{T}}}{\mathbf{1}_{D \times N} \mathbf{H}^{f\text{T}}}. \quad (13)$$

$$\mathbf{H}^f \leftarrow \mathbf{H}^f \circ \frac{\mathbf{W}^{f\text{T}} (\mathbf{X} \circ \tilde{\mathbf{X}}^{\circ -1})}{\mathbf{W}^{f\text{T}} \mathbf{1}_{D \times N}}. \quad (14)$$

Since the columns of $\mathbf{W}^f$ must be unit vectors, they are normalized at the end of each multiplicative update step as shown below,

$$\mathbf{w}^f_k \leftarrow \frac{\mathbf{w}^f_k}{\|\mathbf{w}^f_k\|}. \quad (15)$$

## C. Sparsity Constrained Binary Updates

In order to train $\mathbf{H}^s_j$, $\arg\inf_{\mathbf{H}^s_j} \Phi$ is obtained using the binary updating heuristic proposed in this research. This heuristic directly imposes the $L_0$ norm constraint. The shiftable basis matrices $\mathbf{W}^s_j$ remain fixed throughout the training process.

It should be noted that since $h^s_{j,k}(n) \in \{0,1\}$ and $\|\mathbf{h}^s_j(n)\|_0 \leq L_j$, computing the optimal weight matrix $\mathbf{H}^s_j$ is an

NP-hard problem. Therefore a hill-climbing heuristic is proposed. The component of the total error due to any sample $n$ is given by the following expression,

$$\mathbf{e}(n) = \tilde{\mathbf{x}}(n) - \mathbf{x}(n)$$
$$= \mathbf{W}^f \mathbf{h}^f(n) + \sum_j \mathbf{W}_j^s \mathbf{h}_j^s(n) - \mathbf{x}(n)$$
$$= \mathbf{W}^f \mathbf{h}^f(n) + \mathbf{W}_j^s \mathbf{h}_j^s(n) + \sum_{j' \neq j} \mathbf{W}_{j'}^s \mathbf{h}_{j'}^s(n) - \mathbf{x}(n)$$
$$= \mathbf{W}_j^s \mathbf{h}_j^s(n) - \mathbf{r}_j(n). \quad (16)$$

In the above equality, $\mathbf{r}_j(n)$ is the residual approximation of the $n^{\text{th}}$ sample when ignoring the $j^{\text{th}}$ shiftable load as shown below,

$$\mathbf{r}_j(n) = \mathbf{x}(n) - \mathbf{W}^f \mathbf{h}^f(n) - \sum_{j' \neq j} \mathbf{W}_{j'}^s \mathbf{h}_{j'}^s(n).$$
$$\approx \sum_{k \in \mathcal{S}_j} \mathbf{w}_{j,k}^s h_{j,k}^s(n). \quad (17)$$

The objective function for any given sample $n$ provided earlier in (10) can be expressed as given below,

$$\varphi(n) = \frac{1}{2} \|\mathbf{e}(n)\|_2^2$$
$$= \frac{1}{2} \sum_d e_d^2(n)$$
$$= \frac{1}{2} \sum_d \left( \sum_{k \in \mathcal{S}_j} w_{j,d,k}^s h_{j,k}^s(n) - r_{j,d}(n) \right)^2. \quad (18)$$

In (18), consider any term $d$ of the outer summation. The inner summation is carried out over all columns $k \in \mathcal{S}_j$ of $\mathbf{W}_j^s$ such that $w_{j,k,d}^s \neq 0$. Noting that any such non-zero $w_{j,k,d}^s = p_j$, the error $\varphi(n)$ can be re-expressed in the following manner,

$$\varphi(n) = \frac{1}{2} p_j^2 \sum_{d \in \mathcal{D}_k} \left( \sum_{k \in \mathcal{S}_j} h_{j,k}^s(n) - \frac{r_{j,d}(n)}{p_j} \right)^2$$
$$+ \frac{1}{2} \sum_{d \notin \mathcal{D}_k} r_{j,d}^2(n). \quad (19)$$

The summation in (19) above is carried out over all elements in the set of indices $\mathcal{D}_k$ defined as, $\mathcal{D}_k = \{d | w_{j,d,k}^s \neq 0\}$. The set $\mathcal{D}_k$ can be obtained easily by examining $k^{\text{th}}$ column of $\mathbf{W}_j^s$ and including all indices $d$, that have nonzero entries in that column. After scaling appropriately and ignoring the term in (19) not containing $h_{j,k}^s(n)$, the objective is,

$$\varphi'(n) = \sum_{d \in \mathcal{D}_k} \left( \sum_{k \in \mathcal{S}_j} h_{j,k}^s(n) - r'_{j,d}(n) \right)^2. \quad (20)$$

The proposed algorithm begins with $\mathbf{h}_j^s(n) = \mathbf{0}_{|\mathcal{S}_j|}$, updating it in a stepwise manner, one element in each step. In each step $l$ a new index $k \in \mathcal{S}_j$ is selected and the corresponding $h_{j,k}^s(n)$ updated to 1. The algorithm can be implemented using a separate binary heuristic subroutine. However, the correct arguments need to be passed to the subroutine. The steps to do so are shown below.

$$\mathbf{r}_j(n) \leftarrow \mathbf{x}(n) - \mathbf{W}^f \mathbf{h}^f(n) - \sum_{j' \neq j} \mathbf{W}_{j'}^s \mathbf{h}_{j'}^s(n)$$
$$\mathbf{r}_j(n) \leftarrow \frac{1}{p_j} \mathbf{r}_j(n)$$
$$\mathbf{h}_j^s(n) \leftarrow \texttt{hillClimb}(\mathbf{W}_j^s, \mathbf{r}_j(n))$$

The arguments of subroutine `hillClimb()` are a $D \times 1$ vector $\mathbf{r}$ and either a $D \times S$ binary matrix $\mathbf{W}$, or equivalently the sets of indices $\mathcal{D}_k$ for each $k \in \{1, 2, \dots, S\}$. The subroutine returns an updated $\mathbf{h}$ that minimizes,

$$\varphi(\mathbf{h}|\mathbf{W}, \mathbf{r}) = \sum_d \left( r_d - \sum_{k=1}^S w_{k,d} h_k \right)^2. \quad (21)$$

Alternately, the expression in (21) can be written as,

$$\varphi(\mathbf{h}|\mathcal{D}_1, \dots, \mathcal{D}_D, \mathbf{r}) = \sum_{d \in \mathcal{D}_k} \left( r_d - \sum_{k=1}^S h_k \right)^2 + \frac{1}{2} \sum_{d \notin \mathcal{D}_k} r_d^2, \quad (22)$$

The proposed subroutine is outlined in Fig. 4. The two arguments that are passed to the subroutine `hillClimb()` consist of a basis matrix $\mathbf{W}$, and a residual vector $\mathbf{r}$. If $\mathbf{W} \equiv \mathbf{W}_j^s$, the basis matrix of the $j^{\text{th}}$ shiftable load, then $\mathbf{r} \equiv \mathbf{r}_j(n)$ as shown in (17) but after normalization so that $p_j = 1$. It is assumed that two constants, $S$, the number of columns in $\mathbf{W}$, and $L$, the maximum allowable value of $\|\mathbf{h}\|_0$ are implicitly accessible to the subroutine.

The subroutine `hillClimb()` maintains a set $\mathcal{L}$ of all indices $k$ such that $h_k = 0$. Since the subroutine begins with $\mathbf{h} = \mathbf{0}_S$ (step 1), the set $\mathcal{L}$ is initialized to include all basis vectors in $\mathbf{W}$ (step 2). The quantity $l$ stores the value of $\|\mathbf{h}\|_0$; therefore it is initialized to 1 (step 3). The subroutine maintains a Boolean variable `terminate` to indicate if the termination condition of the algorithm is satisfied; it is initialized to **FALSE** (step 4). Since $\mathbf{W}$ is a sparse binary matrix, the sets $\mathcal{D}_k$ ($k = 1, 2, \dots, S$ are initialized to indicate the elements in column $\mathbf{w}_k$ in $\mathbf{W}$ that contain 1s (step 5).

During each iteration of the **while** loop (step 6), an $h_k$ is updated to unity and the corresponding index $k$ removed from $\mathcal{L}$. In other words, the norm $\|\mathbf{h}\|_0$ is increased by unity per iteration. As a hill-climbing procedure, in each iteration the subroutine picks an index $k$ from $\mathcal{L}$ that lowers the error $\|\mathbf{r} - \mathbf{Wh}\|_2^2$ by the maximum amount. In step 6a, the current error is obtained and stored in the variable $\varphi_0$. In the **for** loop that follows in step 5, for every $k$ in $\mathcal{L}$, the error $\varphi_1(k)$ that would result if the corresponding $h_k$ were to be incremented to unity, is computed. However, before updating the $h_k$, the termination condition is evaluated. The indicator variable `terminate` is set to **TRUE** if $\|\mathbf{h}\|_0$ is equal to $L$, in which case no further updates are possible. The variable becomes **TRUE** also if incrementing any other $h_k$ will only increase the error, that is, if the smallest entry of the vector $\boldsymbol{\varphi}_1$ exceeds $\varphi_0$. The latter situation arises when all elements of the residual $\mathbf{r}$ are less than 0.5. This is shown in step 6e.

If the termination condition is not satisfied, the subroutine proceeds by obtaining the index $k$ that corresponds to the smallest $\varphi_1(k)$ (step 6f) and sets that $h_k$ to 1 (step 6g). Next, the same index $k$ is removed from the set $\mathcal{L}$, in step 6f. The variable $l$ in incremented to indicate the new value of $\|\mathbf{h}\|_0$.

```
hillClimb(W,r)
    1. h ← 0_S
    2. L ← {1,2,...,S}
    3. l ← 0
    4. terminate ← FALSE
    5. for each k ∈ L
        5a. D_k ← {d|w_{k,d} ≠ 0}
      end
    6. while terminate == FALSE
        6a. φ_0 ← Σ_d r_d^2
        6b. for each k ∈ L
            6c. φ_1(k) ← Σ_{d∉D_k} r_d^2 + Σ_{d∈D_k} (r_d − 1)^2
          end
        6d. if min_k φ_1(k) ≥ φ_0 or l > L
            6e. terminate ← TRUE
        6f. else
            6g. k ← argmin_k φ_1(k)
            6h. h_k ← 1
            6i. L ← L\{k}
            6j. l ← l + 1
            6k. for each d ∈ D_k
                6l. r_d ← r_d − 1
              end for
        endif
    end while
```

**Fig. 3.** Subroutine for shiftable load weight updates.

It is clear that the subroutine is a hill-climbing heuristic approach to update the weights of the shiftable load basis vectors. Therefore, upon termination, `hillClimb()` returns the binary vector $\mathbf{h} \equiv \mathbf{h}_j^s(n)$ such that,

$$\mathbf{h} = \underset{\|\mathbf{h}\|_0 \leq L}{\arg\inf} \|\mathbf{r} - \mathbf{W}\mathbf{h}\|_2^2. \quad (23)$$

To the best of the authors' knowledge, there are no established upper bound estimates on the computational complexity of mean field annealing, as has been used in [15]. Although it is reasonable to consider $O(|\mathcal{S}_j|)$ as the per iteration complexity in extracting the largest singular value of a sparse matrix (c.f. [19]), this is only an estimate; SVD algorithms as used in [13] for NMF, do not have strict upper bounds. In contrast, the hill climbing procedure described in this research requires $O(L_0 \ln|\mathcal{D}|)$ steps per iteration and $L_0$ can never exceed the total number of basis vectors $|\mathcal{S}_j|$.

### D. NMF Algorithm

The proposed NMF algorithm follows a block coordinated descent method, with the matrices $\mathbf{W}^f$, $\mathbf{H}^f$, and each $\mathbf{H}_j^s$. Whereas the fixed load matrices are treated as in [3],[4],[5], updating the weight matrices of the shiftable loads, a novel feature of this research, is done by calling the subroutine `hillClimb()` described above multiple times in an iterative fashion. The overall algorithm is outlined in Fig. 4.

For simplicity, the initialization steps are omitted. At first, the matrices $\mathbf{W}^f$, $\mathbf{H}^f$ are updated in accordance with (13) and (14). The updates are carried out in steps 1 and 3, with step 2 being the implementation of the normalization step in (15). It may be noted that the informal notation used in step 2 should, in reality, be implemented iteratively to normalize each column of $\mathbf{W}^f$ separately.

In each iteration of step 4, a sample $n \in \mathcal{N}$ is picked, either randomly or sequentially, so that all shiftable load weights can be updated using the hill-climbing procedure. The appearance of a second level for loop in step 5 is due to the presence of more than a single shiftable load. As before, the inner loop may proceed in a sequential manner, or in any random order of the shiftable loads.

Given sample $n$ and a shiftable load $j$, the residual vector is extracted in the manner shown in (17). This is implemented in step 5a. The residual is normalized in step 5b, with respect to its peak $p_j$. Finally in step 5c, the column $\mathbf{h}_j^s(n)$ of the weight matrix $\mathbf{H}_j^s$ is updated by calling `hillClimb()`.

```
while converged == FALSE do
    1. W^f ← W^f ∘ (X ∘ X̃^{∘−1})H^{fT} / (1_{D×N} H^{fT})
    2. W^f ← W^f / ‖W^f‖
    3. H^f ← H^f ∘ (W^{fT}(X ∘ X̃^{∘−1})) / (W^{fT} 1_{D×N})
    4. for each n ∈ N
        5. for each j (in any order)
            5a. r_j(n) ← x(n) − W^f h^f(n) − Σ_{j'≠j} W_{j'}^s h_{j'}^s(n)
            5b. r_j(n) ← (1/p_j) r_j(n)
            5c. h_j^s(n) ← hillClimb(W_j^s, r_j(n))
        end for
    end for
end while
```

**Fig. 4.** Section of the proposed NMF algorithm containing the main loop. Preceding initialization steps have been omitted.

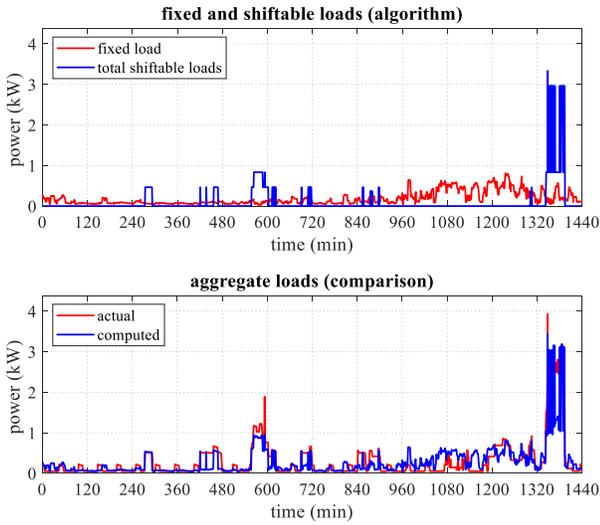

**Fig. 5.** Disaggregated shiftable loads (top) and aggregate loads (bottom) for Day-1.

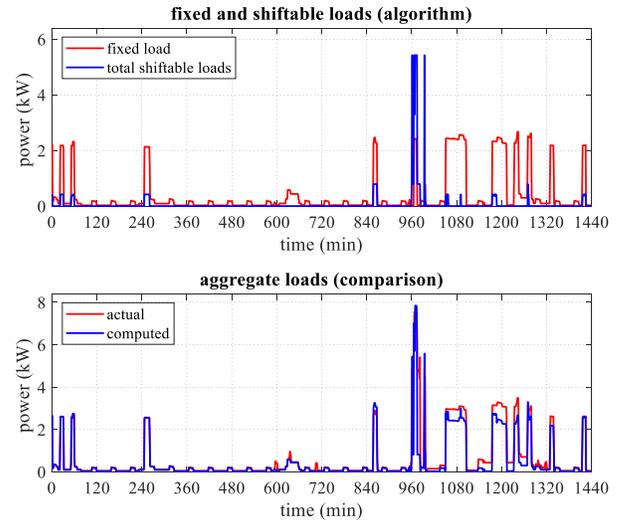

**Fig. 7.** Disaggregated shiftable loads (top) and aggregate loads (bottom) for Day-2.

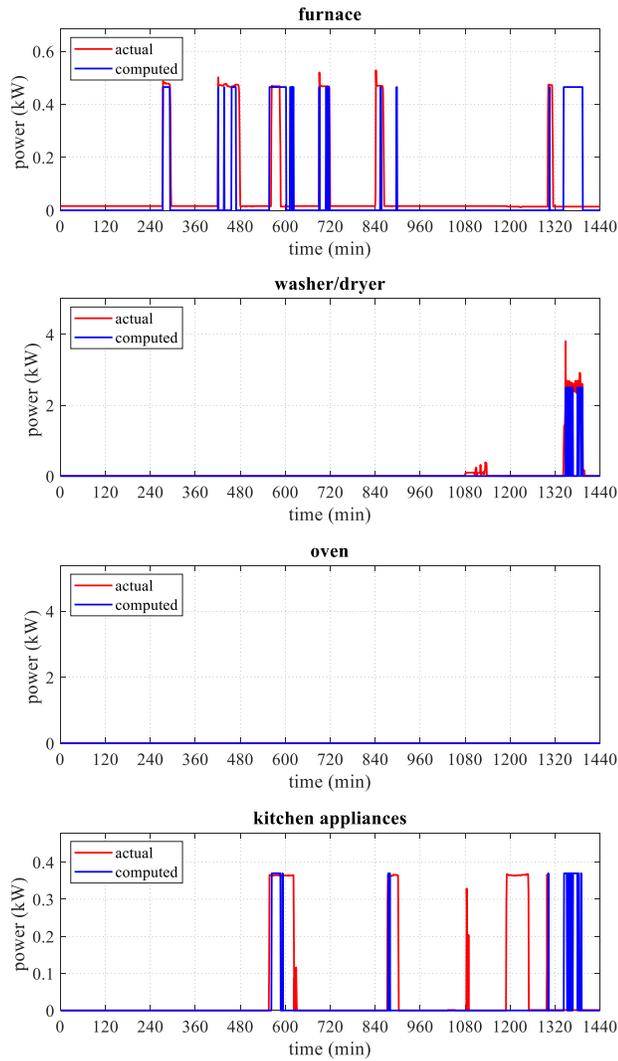

**Fig. 6.** Disaggregated shiftable loads for Day-1.

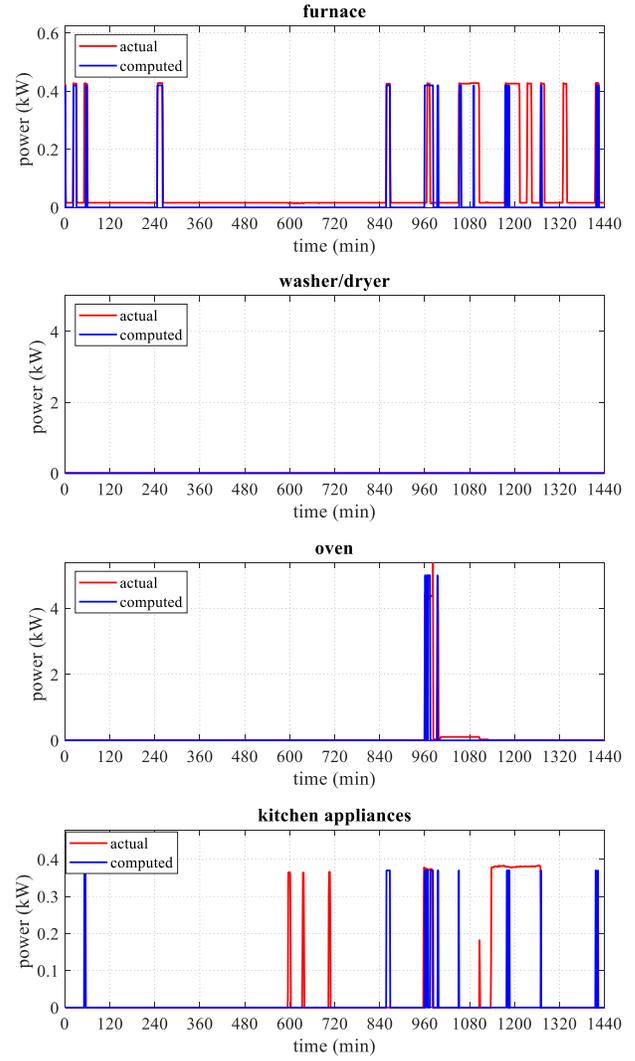

**Fig. 8.** Disaggregated shiftable loads for Day-2.

## IV. RESULTS

### A. Data Preprocessing

The proposed approach was tested on actual energy usage profiles of a single residential customer that was obtained from the Pecan Street Inc., Dataport database [20] sampled at one minute intervals, and for the first three weeks in April, 2019. As energy consumption patterns on weekdays differ significantly from those in weekends, the latter was discarded, yielding a total of $N = 15$ samples. The data was arranged as a $1440 \times 15$ input matrix $\mathbf{X}$ whose columns were the 1440-dimensional sample vectors $\mathbf{x}(n), n \in \mathcal{N}$.

The database in [20] also included individual energy usages of the following four appliances, (*i*) a furnace, (*ii*) a washer/dryer unit, (*iii*) an oven, and (*iv*) kitchen appliances. These measurements provided the basis to evaluate the quality of the disaggregation obtained by the proposed algorithm. All four appliances were classified as shiftable loads for this purpose and indexed in the above order. The rest of the aggregate load was treated as fixed loads.

The peak values $p_j$ of the duty cycle of each load $j \in \{1, 2, 3, 4\}$, as well as the maximum number of its ON cycles, $L_j$, of a typical day (denoted as $L_j^{\max}$) was determined through visual inspection. These parameters are shown in Table I.

TABLE I. DUTY CYCLE PARAMETERS

|  | furnace | washer/dryer | oven | kitchen apps |
|---|---|---|---|---|
| $p_j$ | 0.465 | 2.50 | 5.00 | 0.37 |
| $L_j^{\max}$ | 150 | 20 | 10 | 60 |

### B. Load Disaggregation

The observed $L_j$ maximums (denoted as $L_j^{\max}$) of the loads in Table I served as upper limits on the number of ON cycles, so that constraints of the form $\left\|\mathbf{h}_j^s(n)\right\|_0 \leq L_j^{\max}$ were applied to each shiftable load. The total number of basis vectors were chosen to be $|\mathcal{F}| = 1$ for the fixed loads, and $|\mathcal{S}_j| = 1440$ for each shiftable load $j$. The disaggregated outcomes of the days indexed $n = 1$ and $n = 5$ (indexed after dropping weekends) were picked for illustrative purposes. These will be referred to as hereafter as Day-1 and Day-2.

The plots in Figs. 5 and 6 show the results obtained by the NMF algorithm for Day-1. The disaggregated fixed load $\mathbf{W}^f\mathbf{h}^f(1)$ (red) and the sum of all four shiftable loads, $\sum \mathbf{W}_j^s\mathbf{h}_j^s(1)$ (blue) for the entire 24 hour period appears in Fig. 5. (top). For the sake of comparison, the expected aggregate load $\tilde{\mathbf{x}}(1)$ from the algorithm (blue) is plotted alongside the real load profile $\mathbf{x}(1)$ (red) in Fig. 5 (bottom). The effectiveness of the proposed approach is evident from the similar patterns of both plots. The peak consumption in both cases occur in the late evening hours (1320 – 1440 mins). Additionally the real data shows increased energy usage in the morning hours (540 – 600 mins), which is effectively reproduced by the NMF algorithm.

The disaggregated loads of the individual appliances for Day-1 are provided in Fig. 6. The plots for the furnace, washer/dryer unit, oven, and kitchen appliances appear in order (top through bottom). The actual load data of this 24 hour period appears in red.

Upon close observation, it is apparent that the NMF algorithm accurately reproduces the ON and OFF periods for each load (blue). During Day-1, there were no ON periods for the oven in either case. The ON periods for the washer/dryer unit occur in the interval 1340 – 1400 mins. The effectiveness of the NMF algorithm is obvious from the very strong resemblance of the disaggregated washer/dryer usage to the real data. Likewise, the NMF algorithm faithfully reproduces the real furnace usage profile, albeit to a somewhat lesser extent than before. In comparison to the others, there are some discrepancies in the usage profile that the NMF algorithm yields and the real data for the kitchen appliances.

Figs. 7 and 8 pertain to the energy usage occurring in Day-2. As the results are remarkably similar to those in Day-1, we focus only on a few key observations. In contrast to the previous results, from the actual data in Fig. 7 (bottom) it is seen that the increased usage of shiftable loads during the evening is more spread out. This feature is reflected in the disaggregated signals. Unlike before, there are a few ON cycles for the oven, which is again captured by the algorithm. As in the previous case, it can be seen that the algorithm is not able to pick the ON cycles of the kitchen appliance with the same precision as with other shiftable loads. We attribute the difference to the smaller peak value $p_4 = 0.37$ relative to those of the other appliances.

## V. CONCLUSION

The major contributions of this study are as follows.

(*i*) The load disaggregation of real input data at a significantly high resolution of 1 minute intervals clearly demonstrates the efficacy of the NMF algorithm suggested in this research.

(*ii*) Categorizing individual appliances into fixed and shiftable loads allowed the proposed NMF based algorithm to leverage the observed characteristics of each appliance's ON-cycle.

(*iii*) It has been shown that the NMF algorithm, including the treatment of the shiftable loads, can be viewed in terms of maximum likelihood. This provides a theoretical justification of the new approach introduced in this research.

(*iv*) The hill climbing heuristic, requiring up to only $L_0$ steps per iteration of the outer loop, offers a significant computational advantage over other NMF approaches if used for a similar application. (As an example, this algorithm requires up to $L_0 = L_1^{\max} = 150$ steps for the furnace, whereas an SVD based NMF [13], would need $|\mathcal{S}_1| = 1440$ steps).

In spite of the high fidelity of the disaggregated loads obtained using the approach introduced in this research, there is ample scope for future research. Incorporating a semi-supervised learning algorithm to obtain best fits of the duty cycles would obviate the need of prior knowledge of the duty cycles of individual appliances. Another potential enhancement is to allow the algorithm to learn the ON cycles' probabilities of the appliances. The algorithm may be extended to include these probabilities within the theoretical framework.